\documentclass[aps,pra,twocolumn,superscriptaddress]{revtex4}
\def\be{\begin{equation}}
\def\ee{\end{equation}}
\def\bea{\begin{eqnarray}}          
\def\eea{\end{eqnarray}}
\def\bi{\begin{itemize}}
\def\ei{\end{itemize}}

\usepackage{graphicx}
\usepackage{amsmath}
\usepackage{amssymb}

\begin{document}

\title{Dynamics of the Bose-Hubbard model: transition
from Mott insulator to superfluid}

\author{Fernando M. Cucchietti}
\author{Bogdan Damski}
\affiliation{Theory Division, Los Alamos National Laboratory, Los Alamos, NM 87545, USA}
\author{Jacek Dziarmaga}
\affiliation{Institute of Physics and Centre for Complex Systems, Jagiellonian University,
          Reymonta 4, 30-059 Krak\'ow, Poland}
\author{Wojciech H. Zurek}
\affiliation{Theory Division, Los Alamos National Laboratory, Los Alamos, NM 87545, USA}

\begin{abstract}
We study the dynamics of phase transitions in the 
one dimensional  Bose-Hubbard model. 
To drive the system from Mott insulator to superfluid phase,
we change the tunneling frequency at a finite rate. 
We investigate the build up of correlations during fast and slow 
transitions using variational wave functions, 
dynamical Bogoliubov theory, Kibble-Zurek 
mechanism, and 
numerical simulations. We show that time-dependent correlations
satisfy characteristic scaling relations that can be measured in optical lattices
filled with cold atoms. 
\end{abstract}

%\PACS{} %0
\maketitle

%3.65.-w, 73.43.Nq, 03.75.Lm, 32.80.Bx, 05.70.Fh

\section{Introduction}

The spectacular experimental realization of the Bose-Hubbard  model (BHM)
using cold atoms in an optical lattice \cite{greiner} triggered an avalanche
of both theoretical and experimental activity \cite{zwerger,annals}. The
excitement comes mostly from the fact that the derivation of the BHM in
this system can be carried out rigorously  \cite{zwerger,jaksz}, 
its parameters can be experimentally manipulated in real time \cite{greiner}, 
and lattice geometry can be engineered almost at
will: it can be one, two, three  dimensional, and can have different
shapes, e.g., rectangular, triangular, etc. 

Physics of the Bose-Hubbard model is of both fundamental and practical 
interest. Indeed, the BHM is one of the model systems on which our 
understanding of quantum phase transitions (QPTs) is based
\cite{sachdev,Fisher1989}. The quantum phase transition happens in the BHM 
between the gapless superfluid (SF) phase and the gapped Mott insulator (MI) 
phase. Recently its signatures have been experimentally observed \cite{greiner}. 
In a homogeneous system at fixed density, the transition takes place only when 
the number of atoms is commensurate with the number of lattice sites. The 
practical interest in the BHM originates from the possibility of realization of 
a quantum computer in a system of cold atoms placed in an optical lattice \cite{qc}. 

In spite of experimental studies of the BHM and the large number of numerical 
and analytical contributions, 
understanding of the BHM physics is far from complete. In particular, a theory 
of the dynamics of the MI - SF quantum phase transition is still in its initial 
stages \cite{jacek02,girvin,sengupta,altman}. This is not surprising, as until 
very recently \cite{jacek02,polkovnikov,dorner,jacek,cherng,altman}, QPTs have 
been studied as a purely equilibrium problem. The recent progress in dynamical 
studies has been obtained after applying the Kibble-Zurek mechanism (KZM) 
\cite{kibble,zurek}, which was successful in accounting for non-equilibrium 
aspects of thermodynamical phase transitions \cite{experiment}, to the quantum 
case \cite{dorner,jacek,cherng,bodzio,bodzio1}.

%%%%%%%%%
In this paper we investigate the dynamics of the one dimensional (1D) BHM,
focusing on two-point correlation functions. To describe their time dependence, 
we develop and use a variety of analytical approximations. We find that the 
two-point correlations satisfy simple characteristic scaling relations that 
should be experimentally measurable. Finally, we check the accuracy of our 
predictions with numerical simulations.

Section \ref{model} presents the model and defines the quantities of interest. 
In Section \ref{two} we discuss predictions coming from a toy two-site model. 
Section \ref{fast} (\ref{slow}) analyzes scaling relations of correlation 
functions induced by fast (slow) changes of the tunneling coupling.
%%%%%%%%%

\section{The model}
\label{model}

In terms of
dimensionless variables used throughout this paper, the Hamiltonian reads
\begin{equation}
\label{H}
\hat{H}= -J \sum_{i=1}^M (\hat{a}_{i+1}^\dag \hat{a}_i + {\rm h. c.})
+\frac{1}{2} \sum_{i=1}^M \hat{n}_i(\hat{n}_i-1),
\end{equation}
where we additionally assume a density of one particle per site. Such a model 
should be experimentally accessible in a ring-shaped optical lattice \cite{amico}, 
where the geometry of the problem imposes periodic boundary conditions on (\ref{H}).
Another setup for investigations of the Bose-Hubbard model (\ref{H}) will be provided 
by the ongoing experiment in the Raizen group \cite{raizen}, where a 1D homogeneous 
model with open boundary conditions will be realized. Below, we will assume 
periodic boundary conditions for the sake of convenience, but in a realistic 
experimental situation with a few tens of lattice sites 
the dynamics should be unaffected by the boundary conditions.

The Hamiltonian is driven from the MI to the SF regime
by a linear ramp of the tunneling coupling 
\be
J(t)=\frac{t}{\tau_Q},
\label{Jt}
\ee
where $\tau_Q$ is the quench time-scale \cite{zurek,antunes}. 
The evolution starts at $t=0$ from 
the ground state of (\ref{H}), i.e., $|1,1,\dots\rangle$, where
the numbers denote boson on-site occupations. The evolution stops at 
$t=\tau_QJ_{max}$, where $J_{max}\gg1$. Therefore, the system ends up 
very far away from the critical point separating MI and SF phases: 
$J\approx0.29$ \cite{Kuhner2000}. 
Experimentally, the change of the tunneling
coupling alone can be achieved by proper manipulation of the lattice
potential amplitude, followed by adjustment of the atomic interaction 
strength via Feshbach resonances \cite{eddi}. 

We are interested in the correlation functions:
$$
C_l(t)=\frac{1}{2}\langle\psi(t)|\hat{a}_{i+l}^\dag\hat{a}_i+{\rm h.c.}|\psi(t)\rangle,
$$
which are directly experimentally measurable because momentum distribution of atoms
in a lattice is their Fourier transform $\sim\sum_l\exp(ikl)C_l$ 
($k$ is the atomic momentum) \cite{zwerger}. 
This observation shows that the correlation functions are  
good observables for our problem: by the end of time evolution $J\gg1$ so that 
interactions between atoms are asymptotically negligible. As a result, the correlation 
functions take well defined final values.

\section{Dynamics of two site Bose-Hubbard model}
\label{two}

In this section we consider a toy $2$-site model, a problem that can be completely 
solved analytically. The results of this section will be useful later  for studies of 
larger systems. Using symmetries of the Hamiltonian, one can show that the evolution 
starting from the uniform ``Mott'' state $|1,1\rangle$ leads to 
\begin{equation}
|\psi(t)\rangle= a(t) |1,1\rangle+
b(t)\frac{|2,0\rangle+|0,2\rangle}{\sqrt{2}}, 
\label{ab}
\end{equation}
where $|a|^2+|b|^2=1$ and  
\begin{equation}
i\frac{\partial}{\partial t}
\left(
\begin{array}{c}
a \\
b
\end{array}
\right)=
\left(
\begin{array}{cc}
0 & -2\frac{t}{\tau_Q} \\
-2\frac{t}{\tau_Q} & 1
\end{array}
\right)
\left(
\begin{array}{c}
a \\
b
\end{array}
\right).
\label{almostlz}
\end{equation}
A change of  basis 
\begin{equation}
(a',b')=e^{it/2}(a-b,-a-b)/\sqrt{2}
\label{aprime}
\end{equation}
yields  
\begin{equation}
i\frac{\partial}{\partial t}
\left(
\begin{array}{c}
a' \\
b'
\end{array}
\right)=
\frac{1}{2}
\left(
\begin{array}{cc}
\frac{t}{\tau} & 1 \\
1 & -\frac{t}{\tau}
\end{array}
\right)
\left(
\begin{array}{c}
a' \\
b'
\end{array}
\right) \ \ \ , \ \ \ \tau= \frac{\tau_Q}{4}.
\label{LZtau}
\end{equation}
This is exactly the Landau-Zener (LZ) model \cite{Zener}, whose relevance for dynamics 
of QPTs was recently shown in Refs. \cite{bodzio,bodzio1,dorner,jacek,cherng}.
The quantity of interest is $C_1(t)= 2|b'(t)|^2-1, $where $b'(t)$ is provided by the 
exact solution of the Landau-Zener model in the case when the system starts its time 
evolution from the ground state at $t=0$, i.e., from the anti-crossing center 
\cite{bodzio,bodzio1}. This solution is a superposition of Weber functions 
(see Appendix B of Ref. \cite{bodzio1}), and it leads to 
\begin{eqnarray}
C_1(\infty) &=& -1+\frac{4}{\pi\tau} \sinh\left(\frac{\pi\tau}{4}\right)e^{-\pi\tau/8}
   \left|\Gamma\left(1+\frac{i\tau}{8}\right)+\right. \nonumber\\&&\left.
   e^{i\pi/4}\sqrt{\frac{\tau}{8}}
   \Gamma\left(\frac{1}{2}+\frac{i\tau}{8}\right)\right|^2,
\label{exact}
\end{eqnarray}
which has the following small $\tau_Q$ (fast quench) expansion
\begin{equation}
\label{kinfty}
C_1(\infty)= \frac{\sqrt{\pi}}{4}\sqrt{\tau_Q}+ {\cal O}(\tau_Q^{3/2}).
\end{equation}
For large $\tau_Q$ (slow quench), we expand the gamma functions
for large absolute values of the argument \cite{gradshteyn},
\be
\Gamma (z)= \sqrt{2 \pi} z^{z-\frac{1}{2}}e^z\left( 1 + \frac{1}{12 z}+\frac{1}{288 z^2}
+{\cal O}(z^{-3}) \right), \nonumber
\ee
and use that
\begin{eqnarray}
|\Gamma(i x)|^2 &=& \frac{\pi}{x \sinh(\pi x)}, \nonumber \\
\left|\Gamma\left(\frac{1}{2}+i x\right)\right|^2 &=& \frac{\pi}{\cosh(\pi x)}, \nonumber \\
\left(\frac{1+ i x}{\frac{1}{2}-i x} \right)^{i x} &
\underset{x \rightarrow \infty}{\longrightarrow} &
e^{-\pi x+\frac{3}{2}+\frac{3 i}{8 x}-\frac{3}{8 x^2}}, 
\nonumber
\end{eqnarray}
to obtain
\be
C_1(\infty)= 1- \frac{8}{\tau_Q^2}+{\cal O}(\tau_Q^{-4}).
\label{infty}
\ee
Eq. (\ref{infty}) is surprising since $C_1(\infty)=1-2p_{ex}$, with $p_{ex}$ 
the excitation probability of the LZ system (\ref{LZtau}) at $t=\infty$. Indeed, 
it implies that the excitation probability equals
$$
p_{ex}(t:0 \rightarrow \infty) = \frac{4}{\tau_Q^2}
$$
when the LZ system (\ref{LZtau}) starts evolution from anticrossing center ($t=0$)
and evolves slowly till $t=\infty$, while it is 
{\it exponentially} small (assuming $\tau_Q\gg1$)
$$
p_{ex}(t:-\infty \rightarrow \infty) = \exp\left(-\frac{\pi\tau_Q}{8}\right),
$$ 
for the  LZ model evolving from $t=-\infty$ to $t=\infty$.

We have verified numerically that a slightly larger system ($4$ atoms in $4$ lattice 
sites) exhibits the same scaling of $C_1(+\infty)$ in the fast and slow transition 
limit. Thus, these characteristics are not specific to a 2-site toy system only. In 
the following sections we will use different techniques to argue that the same scaling 
properties are shared by large lattice models.

Before proceeding further, however, we mention that the power-law behavior of 
the excitation probability when the evolution starts at the anticrossing can be 
relevant for quantum adiabatic algorithms \cite{farhi}. Indeed, by starting (or, 
by symmetry \cite{bodzio1}, ending) the algorithm near the anticrossing center
the computation has a much higher failure probability. Thus, such situations have 
to be fiercely avoided when designing a path in Hamiltonian space between
the initial and the solution Hamiltonians.

\section{Fast transitions}
\label{fast}

\begin{figure}[t]
\includegraphics[width=\columnwidth,clip=true]{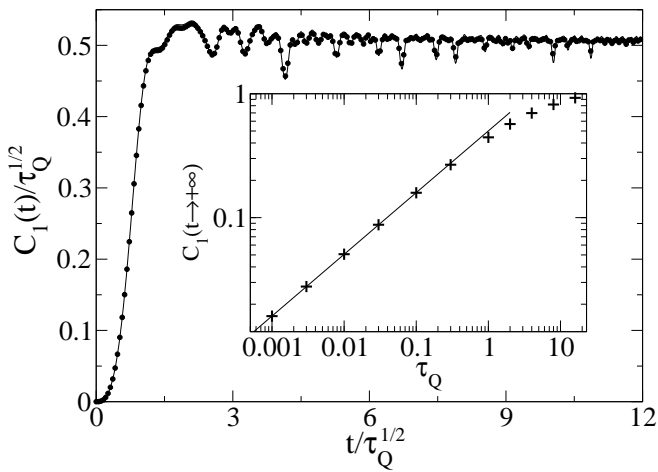}
\caption{Scaling properties of the first correlation function obtained 
         numerically.
         Solid line: $\tau_Q=0.001$, dots: $\tau_Q=0.03$. 
         Inset: solid line is a power law fit to data
	 for $0.001\le\tau_Q\le0.1$ giving $C_1(\infty)=0.501\tau_Q^{0.498}$.
	 All data is for $M=10$ and $J_{max}=600$.}
\label{scaling}
\end{figure}

In this section we consider systems undergoing fast ($\tau_Q\ll1$) quenches. 
Let us start by summarizing some relevant numerical findings on $C_1$.
We studied numerically system sizes $M=3\cdots10$ ($M$ is the number of
lattice sites/atoms), and found that in all cases \cite{inft}
\begin{equation}
C_1(\infty)=\alpha\tau_Q^\beta
\label{beta}
\end{equation}
for $\tau_Q$'s smaller than about $10^{-1}$. Depending on the system size,
$\alpha\in(0.37,0.5)$ while  $\beta$ equals $1/2$ within fitting errors: see
the inset of Fig. \ref{scaling} for the $M=10$ case. 

Moreover, as depicted in Fig. \ref{scaling}, the whole $C_1(t)$ function
after the rescaling 
\begin{equation}
C_1(t)\to \frac{C_1\left(\frac{t}{\sqrt{\tau_Q}}\right)}{\sqrt{\tau_Q}}
\label{sc_}
\end{equation}
takes an universal form for $\tau_Q$ smaller than about $10^{-1}$.

Two remarks are in order now. First, the two site prediction, Eq. 
(\ref{kinfty}), shares the same scaling with $\tau_Q$ and a prefactor of 
the same order of magnitude ($\sqrt{\pi}/4\approx0.44$) as the numerics
for larger systems. Second, it is interesting to ask weather the scaling 
relation (\ref{beta}) can be experimentally verified. 
Taking $10^{-1}$ as  the  largest $\tau_Q$ for which (\ref{beta}) works very well,   
we get $C_1(\infty)\approx0.16$ in $M=10$ case (Fig. \ref{scaling}).
This is to be compared to the ground state predictions  at (i) the critical point 
($C_1\approx0.8$ \cite{DZ}), and (ii) the asymptotic value deep in superfluid ($C_1=1$).
Thus, our results suggest that, despite the fast drive of the system through the transition
point, the first correlation builds up macroscopically. Therefore it should be 
experimentally measurable.

In the following we will explain the observed behavior of $C_1$ 
first by the time-dependent  perturbation theory,
and then by developing a bosonic Bogoliubov theory.

\subsection{Short time diabatic dynamics}
\label{short}

For short times we can approximate the wave function using time-dependent 
perturbation theory,
\begin{eqnarray}
&&|\psi(t)\rangle= a(t)|1,1,\dots\rangle+
b(t)\left(|0,2,1,\dots\rangle+|2,0,1,\dots\rangle \right. \nonumber \\
&& \left. +|1,0,2,1,\dots\rangle+ |1,2,0,1,\dots\rangle+\cdots\right)/\sqrt{2M},
\label{wp}
\end{eqnarray}
where $M>2$ is assumed and $|a|^2+|b|^2=1$. 
A time-dependent variational principle  predicts in this case that dynamics of 
$a(t), b(t)$ is governed by Eq. (\ref{almostlz}) with $\tau_Q$  replaced by
$\tau_Q/\sqrt{M}$. Therefore, the familiar LZ problem shows up again, 
and it is useful to define  new amplitudes $a'$ and $b'$ in the same way as in 
(\ref{aprime}). Dynamics of $a'(t)$ and $b'(t)$ is governed by 
Eq. (\ref{LZtau}) with $\tau\to\tau/\sqrt{M}$. 

\begin{figure}[t]
\includegraphics[width=\columnwidth,clip=true]{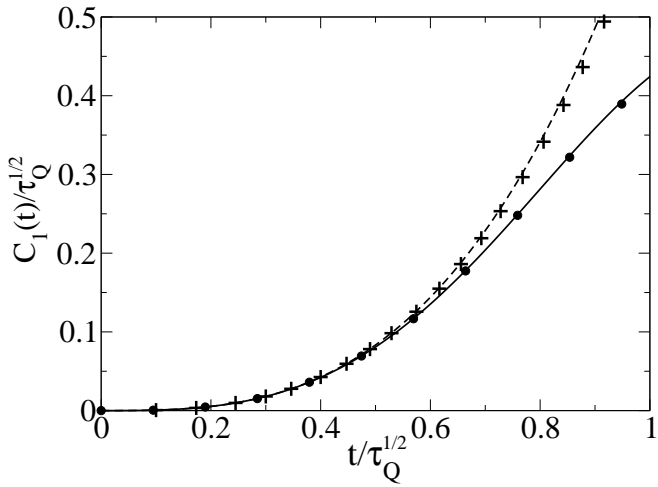}
\caption{Short time dynamics of the first correlation function.
         Numerics for $M=10$ and $J_{max}=600$ is given by 
	 solid line ($\tau_Q=0.001$) and large dots ($\tau_Q=0.1$).
	 The dashed line presents Eq. (\ref{k_t}). The pluses ($+$) stand for 
	   a numerical solution in the Bogoliubov model for 
	   $\tau_Q=0.001$ and prediction (\ref{hatKl}) for $l=0$ (both data
	   overlap).}
\label{k_t_plot}
\end{figure}

To describe the build up of 
$$C_1(t)=(2|b'(t)|^2-1)/\sqrt{M}$$
for the wave-function (\ref{wp}), 
we expand the exact solution of $b'(t)$ \cite{bodzio1} for small $\tau_Q$
obtaining, in the lowest order,
\begin{eqnarray}
\label{k_t}
\frac{C_1(t)}{\sqrt{\tau_Q}}= \frac{2}{3} 
\left[\frac{t}{\sqrt{\tau_Q}}\right]^3.
\end{eqnarray}
Expression (\ref{k_t}) is  interesting: it implies  that the way in which 
the first correlation builds up over time is independent of system size
and takes some universal (independent of $\tau_Q$) form after simple rescalings.
In Fig. \ref{k_t_plot} this prediction is compared to the numerical solution of 
the $10$-site Hubbard model. A perfect agreement is found for times smaller than about 
$\frac{1}{2}\sqrt{\tau_Q}$. As will be explained in Sec. \ref{bogoliubov}, 
the number fluctuations start to develop significantly around 
$t \sim \sqrt{\tau_Q}$, so it is not surprising that a simple wave function (\ref{wp}) 
fails to describe subsequent  dynamics.

\subsection{Bogoliubov theory}
\label{bogoliubov}

Using the insight gained from the above studies, we develop Bogoliubov 
approach that includes a macroscopic number of excitations
into the wave function and is able to describe longer than nearest
neighbor correlations. Our approach is a variant of the theory
developed by Altman and Auerbach \cite{altman} for large density of 
particles.

We truncate the Hilbert space to states with only $\{0,1,2\}$ particles
per site. The initial state is the Mott state with
exactly $1$ particle per site. In a fast transition ($\tau_Q\ll1$)
we can get well into the superfluid regime of $J\gg 1$ before any substantial 
number fluctuations have a chance to build up around the initial Mott state.
Thus, in a fast transition the truncation remains valid well in the
superfluid regime.

As already mentioned, the correlators $C_l$ are conserved by the hopping
term in the Hamiltonian. The hopping term dominates when $J\gg 1$ and this
is why in this regime the correlators are observed to be more or less
constant, see Fig. \ref{scaling}.  Our idea is to use a truncated theory
to predict correlators $C_l(t)$ up to an instant $\tilde t$ so large that
$J(\tilde t\,)\gg1$, but small enough to keep the number fluctuations
negligible. The predicted correlators do not change in the following
evolution dominated by the hopping term, so that $C_l(\tilde t\,)\approx
C_l(\infty)$.

In the truncated Hilbert space we call $2$ particles in a site a
quasiparticle, and an empty site is called a quasihole. The Mott state
with $1$ particle in each site is now the ``empty'' vacuum state.
We introduce the quasiparticle and quasihole creation operators as
$\hat c_i^\dag$ and $\hat d_i^\dag$, respectively. 
Their action is best illustrated by mapping the boson occupation number 
onto two numbers, $(n_c,n_d)$, 
where $n_c$ ($n_d$) are quasiparticle (quasihole) occupation numbers.
This way we have in each site:
$|2\rangle=|(1,0)\rangle$, $|1\rangle=|(0,0)\rangle$, 
and $|0\rangle=|(0,1)\rangle$.
Since within the $\{0,1,2\}$ subspace we cannot have two 
quasiparticles/holes in the same site, the hard-core constraint
has to be implemented: $(\hat c_i^\dag)^2\equiv0$ and $(\hat d_i^\dag)^2\equiv0$.
All this leads to 
$\protect{\hat c_i|...,(n_c,n_d)_i,...\rangle= \delta_{1n_c} 
|...,(n_c-1,n_d)_i,...\rangle}$,
$\protect{\hat c_i^\dag|...,(n_c,n_d)_i,...\rangle= 
\delta_{0n_c} |...,(n_c+1,n_d)_i,...\rangle}$,
and analogical relations for the action of $\hat d_i$ and $\hat
d_i^\dag$ operators.
Additionally, we have to  remove the states with one quasiparticle and
quasihole in the same site, $|...,(1,1)_i,...\rangle$, since these states also 
do not map onto the $\{0,1,2\}$ subspace. This is done by the 
projector $\hat P=\prod_i (1-\hat c_i^\dag\hat c_i\hat d_i^\dag\hat d_i)$. 
Finally, we note that since we deal with hard-core {\it bosons}  all the 
commutators of quasiparticle/hole operators at different lattice sites
commute.

In this new language the Hamiltonian (\ref{H}) in the  $\{0,1,2\}$ subspace 
equals exactly $\hat P \hat{H}_2 \hat P$, where $\hat{H}_2$ is quadratic in 
quasihole/quasiparticle operators
\bea
\hat{H}_2 &=& -J\sum_{\langle i,j \rangle}
\left[
2 \hat c^\dagger_i \hat c_j + \hat d^\dagger_i \hat d_j +
\sqrt{2}
\left(\hat d_i \hat c_j + \hat d_i^\dag c_j^\dag\right)
\right]
\nonumber\\&+&
\sum_i\hat c^\dagger_i \hat c_i,
\label{Ht}
\eea
where $\langle i,j \rangle$ denotes nearest neighbor pairs.
We also mention here that the new  operators satisfy periodic 
boundary conditions as the original bosonic operators do.

The  truncated 
Hamiltonian $\hat P \hat H_2 \hat P$ is exact in the $\{0,1,2\}$ subspace,
but it is not quadratic in $\hat c$ and $\hat d$. In order to proceed we approximate 
$\hat{H}\approx \hat{H}_2$ and lift the hard-core bosonic constraint in all
subsequent calculations:  
from now on $[\hat c_i,\hat c_j^\dag]=\delta_{ij}$ and 
$[\hat d_i,\hat d_j^\dag]=\delta_{ij}$. 
This way we arrive at a bosonic theory with a quadratic 
Hamiltonian $\hat H_2$ leading to solvable linearized equations of motion.
The quadratic theory remains self-consistent as long as average density of 
excitations \cite{remark}
\be
\rho_{\rm ex}=
\langle  \hat c_i^\dagger \hat c_i \rangle =
\langle  \hat d_i^\dagger \hat d_i \rangle \ll \frac12.
\label{dens}
\ee
When $\rho_{\rm ex}\gtrsim \frac12$ it is likely to find 
quasiparticles and quasiholes occupying the same lattice site and the constraint 
imposed by the projector $\hat P$ must be violated.
 
We  proceed by going to the momentum space 
$$\hat c_r=\frac{1}{\sqrt{M}}\sum_k\hat c_k e^{ikr}, 
\ \hat d_r=\frac{1}{\sqrt{M}}\sum_k\hat d_k e^{ikr}.$$
To simplify time-dependent calculations we add to the Fourier transformed 
Hamiltonian two terms, $\sum_k J(\hat c_k^\dag\hat c_k-\hat d_k^\dag\hat d_k)\cos
k$ and $\sum_k \frac{1}{2}(\hat d_k^\dag\hat d_k-\hat c_k^\dag\hat c_k)$,
that both commute with the Hamiltonian itself and therefore do not change 
the evolutions considered in this paper. The resulting Hamiltonian becomes
\bea
\hat{H}_2&=& -J\sum_k 
\cos k
\left[3\hat c^\dagger_k \hat c_k + 3\hat d^\dagger_k \hat d_k +
      2\sqrt{2}(\hat c_k \hat d_{-k}+{\rm h.c.})
\right]
\nonumber\\
&+&
\frac12\sum_k
\left(
\hat c^\dagger_k \hat c_k + \hat d^\dagger_k \hat d_k
\right).
\label{Hk}
\eea
It can be conveniently rewritten to the form
\begin{eqnarray}
\hat H_2&=&\sum_k[\hat c_k^\dag,-\hat d_{-k}] 
\left[
\begin{array}{cc}
\frac{1}{2}-3J\cos k & -2\sqrt{2}J\cos k \\
2\sqrt{2}J\cos k  & 3J\cos k-\frac{1}{2}
\end{array}
\right]
\left[
\begin{array}{c}
\hat c_k \\ \hat d_{-k}^\dag 
\end{array}
\right]\nonumber\\ 
&+&\sum_k\left(3J\cos k-\frac{1}{2}\right).
\label{H22}
\end{eqnarray}

Below we look at description of time evolutions leaving the discussion of
static properties of our theory to Appendix \ref{appendix}. 
As in former sections we start time evolution from $J=0$,
and $J(t)$ is given by (\ref{Jt}).
We work in the Heisenberg picture where
the system wave function (the ground state at $J=0$: $|1,1,...\rangle$) 
is time-independent while the operators evolve according to 
$i\frac{d}{dt}\hat c_k=[\hat c_k,\hat H_2]$ 
and $i\frac{d}{dt}\hat d_k=[\hat d_k,\hat H_2]$. It leads to 
\begin{equation}
i\frac{d}{dt}\left[
\begin{array}{c}
\hat c_k \\ \hat d_{-k}^\dag  
\end{array}\right]=
\left[
\begin{array}{cc}
\frac{1}{2}-3\frac{t}{\tau_Q}\cos k & -2\sqrt{2}\frac{t}{\tau_Q}\cos k \\
2\sqrt{2}\frac{t}{\tau_Q}\cos k  & 3\frac{t}{\tau_Q}\cos k-\frac{1}{2}
\end{array}
\right]
\left[
\begin{array}{c}
\hat c_k \\ \hat d_{-k}^\dag  
\end{array}\right],
\label{cd_time}
\end{equation}
which has the following general solution:
\begin{eqnarray}
\hat c_k(t) &=& u_k(t)\hat c_k(0)+ \tilde v_k(t)\hat
d^\dag_{-k}(0),\nonumber\\
\hat d^{\dag}_{-k}(t)&=& v_k(t)\hat c_k(0)+ \tilde u_k(t)\hat
d^\dag_{-k}(0),\nonumber
\end{eqnarray}
with initial conditions $u_k(0)=\tilde u_k(0)=1$, $v_k(0)=\tilde v_k(0)=0$. 
After some algebra based on (\ref{cd_time}) one finds that 
$\tilde v_k= v_k^*$ and $\tilde u_k= u_k^*$ -- this simplification showed up
thanks to the convenient addition of the two constants of motion to the 
Fourier transformed quadratic Hamiltonian (see above).
The time evolution of the  modes is given by
\begin{equation}
i\frac{d}{dt}\left[
\begin{array}{c}
u_k \\ v_k  
\end{array}\right]=
\left[
\begin{array}{cc}
\frac{1}{2}-3\frac{t}{\tau_Q}\cos k & -2\sqrt{2}\frac{t}{\tau_Q}\cos k \\
2\sqrt{2}\frac{t}{\tau_Q}\cos k  & 3\frac{t}{\tau_Q}\cos k-\frac{1}{2}
\end{array}
\right]
\left[
\begin{array}{c}
u_k \\ v_k  
\end{array}\right].
\label{uv_time}
\end{equation}
Additionally, we see that 
the Bose commutation between the time-dependent operators 
requires that $|u_k(t)|^2-|v_k(t)|^2=1$, which is conserved by the time
evolution (\ref{uv_time}). All expectation values can be 
calculated after solving (\ref{uv_time}) using the fact that 
the wave-function in the Heisenberg picture
is $|\Psi\rangle=|1,1,...\rangle$ for all times,
so that  $\hat d_r(0)|\Psi\rangle= 0$ and $\hat c_r(0)|\Psi\rangle= 0$.

In the following we use perturbative solution of Eqs. (\ref{uv_time}) in powers
of $\sqrt{\tau_Q}$. The discussion is simplified by introducing a new time-like
variable
\be
s =\frac{t^2}{\tau_Q},
\ee
whose form is motivated by the scaling property (\ref{sc_}). Equations 
(\ref{uv_time}) become
\bea
&&
i\frac{d}{ds}
\left[
\begin{array}{c}
u_k \\ v_k 
\end{array}\right]+
\cos k
\left[\begin{array}{cc}
 \frac32  &  \sqrt{2}  \\
-\sqrt{2} & -\frac32 
\end{array}\right]
\left[
\begin{array}{c}
u_k \\ v_k 
\end{array}\right]= \nonumber\\
&&
\frac{\sqrt{\tau_Q}}{4\sqrt{s}}
\left[\begin{array}{cc}
1  &   0  \\
0  &  -1 
\end{array}\right]
\left[
\begin{array}{c}
u_k \\ v_k 
\end{array}\right],
\label{uv_s}
\eea
with $u_k(0)=1$ and $v_k(0)=0$. These equations can be solved iteratively 
in powers of the small parameter $\sqrt{\tau_Q}\ll1$.

As a self-consistency check, we calculate the density of excitations, 
Eq. (\ref{dens}). Assuming fast transition limit, 
$\tau_Q\ll1$, we solve Eqs. (\ref{uv_s}) to zero order in $\sqrt{\tau_Q}$ 
and find that
$$
\rho_{ex}~=~
\frac{1}{2\pi}
\int_{-\pi}^\pi dk |v_k|^2 \approx s^2 
$$ 
for small $s$. Therefore, $\rho_{ex}\ll \frac12$ for $s\ll \frac{1}{\sqrt{2}}$ so 
that the quadratic approximation breaks down at
\begin{equation}
\frac{{\tilde t}\,^2}{\tau_Q}
~\equiv~
\tilde s
~\simeq~
\frac{1}{\sqrt{2}}~,
\label{O}
\end{equation}
or at $\tilde t\simeq \sqrt{\tau_Q/\sqrt{2}}$. In a linear quench (\ref{Jt}) 
this break-down time corresponds to
$$
\tilde J \simeq \frac{1}{\sqrt{\sqrt{2}\tau_Q}}\gg 1
$$
which is well in the superfluid regime for a fast transition. 
Therefore, when $\tau_Q\ll 1$, our linearized Bogoliubov approach 
does not break down until well in the superfluid regime. 

These calculations prove that Bogoliubov approach works reliably before 
$\tilde J\gg 1$ and the correlation functions  are (see Appendix \ref{appendix}
for static predictions) \cite{cl}
\be
C_l= 
\int_{-\pi}^\pi \frac{dk}{2\pi}~
\cos(kl) 
\left[ 
3|v_k|^2+\sqrt{2}(u_kv_k^*+u_k^*v_k)
\right].
\label{Cl}
\ee
Solving (\ref{uv_s}) we find that 
$C_{2l}(t)={\cal O}(\tau_Q)$, while 
\be
\frac{C_{2l+1}(t)}{\sqrt{\tau_Q}}=
\frac{8\pi~s^{3/2}}{3}~
\sum_{n=l}^\infty
(-1)^{n+1}~
\alpha_{l,n}~
s^{2n}~,
\label{hatKl}
\ee
with coefficients
\be
\alpha_{l,n}=
\frac{\left(\frac34\right)_n
      \Gamma(2+2n)
      \Gamma\left[1-2l+2n\right]^{-1}
      \Gamma\left[3+2l+2n\right]^{-1}}
     {n!
      \left(\frac32\right)_n
      \left(\frac74\right)_n 
      \Gamma\left[-\frac12-l-n\right]
      \Gamma\left[\frac12+l-n\right]},
\nonumber
\ee
$(x)_n\equiv\Gamma[x+n]/\Gamma[x]$. To obtain this series expansion 
we differentiate $\frac{d}{ds} C_l$ in Eq. (\ref{Cl}),
remove the resulting $s$-derivatives with the help of equations of motion 
(\ref{uv_s}), keep only the leading terms $\sim\sqrt{\tau_Q}$, and finally 
integrate such obtained $\frac{d}{ds} C_l$ over $s$ to get $C_l(s)$. 

The first term in the $l=0$ version of (\ref{hatKl})  reproduces Eq.
(\ref{k_t}). As shown in Fig. \ref{k_t_plot}, Eq. (\ref{hatKl}) works
perfectly until $s\simeq 1/\sqrt{2}$, i.e., up to the expected breakdown of the
Bogoliubov approach (\ref{O}).  

Since we consider $\tau_Q\ll1$, we have $\tilde J=\tilde t/\tau_Q\gg 1$, 
and thus the rest of evolution after
$\tilde J$ is dominated by the hopping term that does not change the
correlation functions. Therefore, the correlators at the break-down time
$\tilde t$ are good estimates of the final correlation function:
\begin{equation}
C_l(\infty)\approx C_l(\tilde t\,).
\label{as}
\end{equation}

Setting  $s=\tilde s=1/\sqrt{2}$ in (\ref{hatKl}) for definiteness we get
with accuracy of $O(\tau_Q^{3/2})$
\begin{eqnarray}
C_1(\infty)&\approx& 3.9\times10^{-1}\sqrt{\tau_Q},\nonumber\\
C_3(\infty)&\approx&-3.5\times10^{-3}\sqrt{\tau_Q},  \nonumber\\
C_5(\infty)&\approx& 1.4\times10^{-5}\sqrt{\tau_Q} \nonumber.
\end{eqnarray}
By solving Eqs.(\ref{uv_s}) perturbatively we find with the accuracy 
$O(\tau_Q^2)$ that 
\begin{eqnarray}
C_2(\infty)&\approx& 4.0\times10^{-2}\tau_Q, \nonumber\\
C_4(\infty)&\approx& -3.2\times10^{-4}\tau_Q,\nonumber\\
C_6(\infty)&\approx& 1.1\times10^{-6}\tau_Q.\nonumber
\end{eqnarray}
The first correlation $C_1(\infty)$ fits well our numerical results 
-- compare to (\ref{beta}). Reliable numerical verification of longer
range correlations would require calculations done on systems larger than our $M\le10$.
Indeed, in the small size numerics it is hard to filter out finite size effects 
especially when the long range correlations, which are small in magnitude, are considered.
Nevertheless, the Bogoliubov theory and our finite size numerics agree
that correlations $C_l$ decay fast with the distance $l$.

In contrast to the simple $2$-site toy model of Sec. \ref{two}, the
Bogoliubov approach is able to describe not only nearest-neighbor but also
longer range correlations. However, it turns out that in fast transitions
the correlation functions are dominated by the nearest-neighbor term $C_1$
with other terms being relatively small, if not negligible. This explains 
why already the simple $2$-site toy model gives such surprisingly accurate 
predictions for larger systems in the fast transition limit.

\section{Slow transitions}
\label{slow}

In this section we focus on the limit of slow transitions, i.e.,
$\tau_Q\gg1$.  Numerical studies in this regime are extremely time
consuming, therefore we concentrate only on analytical results. Our
predictions are based on the Kibble-Zurek mechanism that was successful in
describing non-equilibrium thermodynamical phase transitions, and
apparently works for quantum phase transitions as well
\cite{dorner,jacek,cherng,bodzio,bodzio1}.

According to KZM,
excitations of the system after a {\it slow} transition have the characteristic
length-scale \cite{antunes}
\begin{equation} 
\xi~\sim~\tau_Q^{\frac{\nu}{z\nu+1}}~,
\label{xiKZM}
\end{equation}
where $z$ and $\nu$ are critical exponents and the quench time $\tau_Q$ is
taken as $(dJ/dt)^{-1}$ at the critical point ($J$ is the
parameter driving the transition).
For the Bose-Hubbard model the dynamical exponent $z=1$. The MI - SF
transition (at fixed integer density of atoms) in a $d$-dimensional
Bose-Hubbard model belongs to the universality class of a
$(d+1)$-dimensional $XY$ spin model \cite{Fisher1989}.  In one dimension
this mapping implies that $\nu\to\infty$ (Kosterlitz-Thouless transition).
As a result, $1-C_1$, which is proportional to the hopping energy of long
wavelength excitations, should scale for $\tau_Q\gg 1$ as
\begin{equation}
1-C_1(\infty)\sim\xi^{-2}\sim\frac{1}{\tau_Q^2}.
\label{kinetic}
\end{equation}
The exponent $-2$ means a rather steep dependence of the hopping energy on
the quench time $\tau_Q$, which should make it easily discernible
experimentally.

Using (\ref{xiKZM}) and (\ref{kinetic}) it is easy to provide predictions for
two ($\nu\approx0.67$ \cite{Campostrini2001}) and three ($\nu=1/2$ \cite{Fisher1989}) 
dimensional Bose-Hubbard models. In the two dimensional case one has 
$$1-C_1(\infty) \sim \frac{1}{\tau_Q^{0.8}},$$ 
while in the three dimensional model 
$$1-C_1(\infty) \sim \frac{1}{\tau_Q^{2/3}}.$$
It would be very interesting to verify scaling relations shown in this section 
either experimentally or numerically. 

\section{Summary}
\label{summary}

We described build-up of correlations in the BHM during transitions from
Mott insulator to superfluid regime using:  a variational wave function,
the dynamical one dimensional Bogoliubov theory, the Kibble-Zurek
mechanism, and numerical simulations.  The time-dependent correlations
satisfy characteristic scaling relations that are directly experimentally
measurable.

This research was supported by US Department of Energy and NSA. J.D. was
supported in part by Polish Government scientific funds (2005-2008).

\appendix
\section{Ground state properties of  Bose-Hubbard model well in the Mott regime}
\label{appendix}

Here we discuss the ground state properties of the Bose-Hubbard model
predicted by the Bogolubov theory. The Hamiltonian (\ref{Hk}) can 
be diagonalized by the Bogoliubov transformation
\begin{equation}
\hat c_k = u_k \hat B_k + v^*_k \hat A^\dagger_{-k}, \ \
\hat d_{-k}^\dag = v_k \hat B_k + u^*_k \hat A^\dag_{-k},
\label{uv}
\end{equation}
where $u_k(J)$ and $v_k(J)$ determine static properties of the Bogolubov
vacuum. 

Here the Bogoliubov modes $(u_k,v_k)$ are the eigenmodes
of the stationary Bogoliubov-de Gennes equations
\bea
\omega_k
\left[
\begin{array}{c}
u_k \\ v_k 
\end{array}\right]=
\left[\begin{array}{cc}
-3J\cos k+\frac12  & -2\sqrt{2}J\cos k \\
 2\sqrt{2}J\cos k  &  3J\cos k-\frac12 
\end{array}\right]
\left[
\begin{array}{c}
u_k \\ v_k 
\end{array}\right],
\label{BdGstat}
\eea
with positive norm: $|u_k|^2-|v_k|^2=1$ and eigenfrequency
$$
\omega_k= \frac{1}{2}\sqrt{4J^2\cos^2 k-12J\cos k+1}.
$$
The normalization condition guarantees proper, i.e.,
bosonic commutation relations of $\hat A_k$ and $\hat B_k$ operators:
$[\hat A_k,\hat A_p^\dagger]=[\hat B_k,\hat B_p^\dagger]=\delta_{kp}$,
$[\hat A_k, \hat B_p]=0$, etc.
The diagonalized Hamiltonian 
\be
\hat{H}_2=\sum_k \omega_k 
\left(
\hat A^\dagger_k \hat A_k +
\hat B^\dagger_k \hat B_k
\right)+\sum_k \left(\omega_k+3J\cos k-\frac{1}{2}\right)
\label{Hstat}
\ee
is a sum of Bogoliubov quasiparticle excitations. Its ground state is a Bogoliubov
vacuum annihilated by all $\hat A_k$ and $\hat B_k$. 

Now we calculate different quantities assuming that the system size
$M\to\infty$.
As a self-consistency check we calculate the density of excitations
\be
\rho_{\rm ex}=
\frac{1}{2\pi}\int_{-\pi}^\pi dk |v_k|^2 = 4J^2+ O(J^4).
\ee
The density remains  $\ll\frac12$ for $J\ll \frac{1}{2\sqrt{2}}$ -- compare to
(\ref{dens}).

The expression for correlation functions
in the static calculations is obtained after using \cite{cl}.
Due to similarity in the notation, it is the same as (\ref{Cl}), except that 
now $u_k$ and $v_k$ depend on $J$ rather than $t$. As a result we get
\begin{eqnarray}
C_1 &=& 4J   + O(J^3), \nonumber\\ 
C_2 &=&18J^2    + O(J^4), \nonumber\\
C_3 &=& 88J^3   +O(J^5),\nonumber\\  
C_4 &=& 450J^4   + O(J^6), \nonumber\\
C_5 &=& 2364J^5 + O(J^7),\nonumber\\
C_6 &=& 12642J^6 + O(J^8), \nonumber\\
C_7 &=& 68464J^7 + O(J^9). \nonumber
\end{eqnarray}
These results in the {\it lowest} nontrivial order listed above
agree perfectly with perturbative ones. Indeed, $C_1\cdots C_3$ 
can be found in Eq. (5) of  \cite{DZ}, while $C_4\cdots C_7$ match 
unpublished results of one of us (B.D.).
Discussion of this intriguing finding is beyond the scope of this study and 
is left out for further detailed investigations.

To close discussion on static properties of our theory
we notice that also the ground state enenergy per site (${\cal E}$),
predicted by the Bogolubov theory, agrees in the lowest order
with exact perturbative calculation. Indeed, we get from (\ref{Hstat}) that 
$$
{\cal E}= 
\frac{1}{2\pi}\int_{-\pi}^\pi dk \left(\omega_k+3J\cos k-\frac{1}{2}\right)=
-4J^2 + O(J^4),
$$
which has to be compared to Eq. (3) of \cite{DZ}.


\begin{thebibliography}{99}
\bibitem{inft} We approximate in numerical calculations 
$C_1(\infty)$ by $C_1(t=J_{max}\tau_Q)$,
which should be a reasonable assumption as long as $J_{max}\gg1$. In
our calculations $J_{max}=600$.
\bibitem{remark} In our theory $\hat c_i^\dag\hat c_i$
and $\hat d_i^\dag\hat d_i$ are expressed in terms of bosonic 
number operators as  $\frac{1}{2}\hat n_i(\hat n_i-1)$ and 
$\frac{1}{2}(\hat n_i-1)(\hat n_i-2)$, respectively.
Therefore one gets $\langle\hat c_i^\dag\hat c_i\rangle= \langle\hat d_i^\dag\hat d_i\rangle$
as long as $\langle\hat n_i\rangle=1$, which is assumed in Eq. (\ref{dens}).
\bibitem{cl} To calculate $C_l$ one finds that 
within the approximations used in this manuscript
$\hat a_{i+l}^\dag\hat a_i\approx 2\hat c^\dag_{i+l}\hat c_i+ \hat d_{i+l}\hat d^\dag_i
+\sqrt{2}(\hat d_{i+l}\hat c_i+\hat d^\dag_i\hat c^\dag_{i+l})$.
\end{thebibliography}
\end{document}